\documentclass{ifacconf}

\usepackage{graphicx}      % include this line if your document contains figures
\usepackage{natbib}        % required for bibliography

\usepackage{amsmath}
\usepackage{amssymb}

\usepackage{setspace}
\usepackage[flushleft]{threeparttable}

\newcommand{\method}[1]{{\texttt{#1}}}

\begin{document}
\begin{frontmatter}

\title{Runtime Monitoring and Fault Detection for Neural Network-Controlled Systems\thanksref{footnoteinfo}} 

\thanks[footnoteinfo]{This work was supported in part by the Leverhulme Trust Early Career Fellowship under Award ECF-2021-517, and in part by the UK Royal Society International Exchanges Programme under Award IES$\backslash$R3$\backslash$223168. Xianxian Zhao was supported by the SEAI (Sustainable Energy Authority of Ireland) under RD\&D Award 22$\backslash$RDD$\backslash$776.}

\author[First]{Jianglin Lan}, 
\author[Second]{Siyuan Zhan},
\author[Third]{Ron Patton}, 
\author[Fourth]{Xianxian Zhao}

\address[First]{James Watt School of Engineering, 
	University of Glasgow, Glasgow G12 8QQ, UK (e-mail: Jianglin.Lan@glasgow.ac.uk).}
\address[Second]{Department of Mechanical, Manufacturing and Biomedical Engineering, Trinity College Dublin, Dublin 2, Ireland (e-mail: zhans@tcd.ie).}	
\address[Third]{School of Engineering, University of Hull, Hull HU6 7RX, UK (e-mail: R.J.Patton@hull.ac.uk).}
\address[Fourth]{School of Electrical and Electronic Engineering, University College Dublin, Belfield,
	D04 V1W8 Dublin, Ireland (xianxian.zhao@ucd.ie).}

\begin{abstract}
	There is an emerging trend in applying deep learning methods to control complex nonlinear systems. This paper considers enhancing the runtime safety of nonlinear systems controlled by neural networks in the presence of disturbance and measurement noise. A robustly stable interval observer is designed to generate sound and precise lower and upper bounds for the neural network, nonlinear function, and system state. The obtained interval is utilised to monitor the real-time system safety and detect faults in the system's outputs or actuators. An adaptive cruise control vehicular system is simulated to demonstrate effectiveness of the proposed design.

\end{abstract}


\end{frontmatter}
%===============================================================================

\section{Introduction}\label{sec:intro}

Machine learning techniques including deep neural networks (NNs) are powerful for modelling and controlling complex nonlinear systems such as robots and autonomous vehicles \citep{Moe+18,Tang+22}. However, NNs are vulnerable to input perturbations such as noise and adversarial attacks. This is even more problematic when NNs are used to generate real-time control actions for automatic systems such as aircraft \citep{JulianKochenderfer21}, because uncertainties (or deviations) in the NN will be propagated and accumulated  the closed-loop, leading to degraded performance and safety concerns \citep{Bensalem+23}. It is thus important to assure real-time safety of NN-controlled systems. 

Safety assurance for NN-controlled autonomous systems has been looked at from different angles in the literature. Lots of research has been devoted to formal methods for verifying the robustness of NNs against perturbations \citep{Liu+21}. The formal methods are normally based on interval bound propagation and the solving of optimisation problems such as mixed-integer linear programming (MILP) \citep{LomuscioMaganti17}, semidefinite programming (SDP)
\citep{lan+23i}, and linear programming (LP) \citep{Bunel+20}. However, these works consider NNs as stand-alone components for image classification, natural language processing, etc. Some researchers have considered verifying safety of the NN-controlled systems through reachability analysis. Given an initial state set, they use methods, such as MILP \citep{KargLucia20}, SDP \citep{Hu+20}, LP \citep{Everett+21}, or constrained zonotopes \citep{ZhangXu22}, to compute the reachable set which contains all the possible future system trajectories and check whether this set is within the safe region or not. However, reachability analysis is computationally expensive and has been conducted offline. This method is thus unable to account for system changes due to external disturbances and measurement noise or detect system security issues in a timely manner. 

This paper is dedicated to runtime monitoring of NN-controlled autonomous systems to assure their safety. \cite{Cofer+20} propose a rule-based method for monitoring NN-based aircraft taxiing, using a set of monitors to measure the aircraft's relative position to the runway and NN's deviation from its training dataset. 
\cite{Xiang21} and \cite{Wang+23o} develop observers to monitor the state of continuous-time NN-controlled nonlinear systems by taking inspiration from the interval observer technique \citep{efimov2013interval}. 
%The observer consisting of two dynamic models generates the interval bound of the real-time system state trajectory, with the gains being solved from linear matrix inequality problems. 
However, their works do not 
consider system disturbance or measurement noise, and assume the existence of an interval for the nonlinear function without providing a way to derive it. They adopt the auxiliary network (AN) method to compute the output bounds of NNs, which is applicable to a wide range of activation functions including the most widely used Rectified Linear Unit (\method{ReLU}) and \method{tanh} \citep{Dubey+22}, but its conservatism results in loose intervals of the system trajectories.

This paper proposes a new interval observer for discrete-time nonlinear systems under disturbance and measurement noise, where the controller is a deep NN with \method{ReLU} activations. 
The proposed observer design is advantageous over the existing methods \citep{Xiang21,Wang+23o} in three aspects: (i) it uses an MILP optimisation approach to obtain the tightest NN bounds, avoiding the conservatism introduced by the AN method.
(ii) it provides a systematic way to construct the upper and lower bounds of the nonlinear function; (iii) it ensures that the generated state interval is robust against external disturbance and measurement noise.   

%The rest of the paper is structured as follows: Section \ref{sec:problemsetup} describes the problem and some preliminaries, Section \ref{sec:observer} presents the interval observer design, Section \ref{sec:application} describes application of the observer to safety monitoring and fault detection, Section \ref{sec:simulation} reports the simulation results and Section \ref{sec:conclusion} draws the conclusions. 

\section{Problem description and preliminaries}\label{sec:problemsetup}

Consider a discrete-time system modelled
by
\begin{subequations}\label{eq:sys dyn}	
\begin{align}
x_{t+1} &= A x_t + B u_t + F g(x_t) + w_t, \\
y_t &= C x_t + v_t,
\end{align}	
\end{subequations}
where $t$ is the sampling time. $x_t \in \mathbb{R}^n$, $u_t \in \mathbb{R}^m$, $w_t \in \mathbb{R}^n$, 
$y_t \in \mathbb{R}^p$, and $v_t \in \mathbb{R}^p$ are the vectors 
of system state, control inputs, external disturbances, 
measured outputs, and noise, respectively. 
$g(x_t) \in \mathbb{R}^s$ is the system nonlinearity.  
$A \in \mathbb{R}^{n \times n}$, $B \in \mathbb{R}^{n \times m}$, $F \in \mathbb{R}^{n \times s}$, and $C \in \mathbb{R}^{p \times n}$ are known constant matrices. 
In this paper, $u_t = f_\text{nn}(r_t, y_t^o)$, where $r_t \in \mathbb{R}^q$ is the vector of known reference signals and $y_t^o = C_o y_t$ which takes partial $y_t$ using $C_o = [\mathbf{0}_{\ell \times p_1} ~ I_\ell ~ \mathbf{0}_{\ell \times p - \ell - p_1}]$ with the integers $\ell \geq 1$ and $p_1 \geq 0$.
$f_\text{nn} \in \mathbb{R}^m$ is a $L$-layer feedforward NN given by:
\begin{align}
z_0 &= [r_t; y_t^o], ~
z_i = \phi(W_i z_{i-1} + b_i), ~ i \in [1,L-1], \nonumber\\
f_\text{nn}(z_0) &= W_L z_{L-1} + b_L,	\nonumber
\end{align}
where $W_i$ and $b_i$ are the weight and bias of the $i$-th hidden
layer, respectively. $\phi(\cdot)$ is the \method{ReLU}
activation function with the form $\phi(\hat{z}_i) = \max(\hat{z}_i,0)$, where $\hat{z}_i = W_i z_{i-1} + b_i$ and the operation $\max$ is applied elementwise. 

This paper aims to design a robustly stable observer that can generate a precise interval for the state trajectory for runtime safety monitoring and detection of actuator or sensor faults.  

In this paper, the following notations are defined for a matrix $X$: 
$X^+ = \max\{0,X\}$, $X^- = X^+ - X$, and $|X| = X^+ + X^-$. 
The observer design is based on Assumption \ref{assume:signal bounds} and will use Lemma~\ref{lemma:M interval}. 
\begin{assum}\label{assume:signal bounds}
	The pair $(A,C)$ is observable.	
	$x_0 \in [\underline{x}_0, \bar{x}_0]$ for some known constant vector $\underline{x}_0$ and $\bar{x}_0$. 
	$w_t \in [\underline{w}_t, \bar{w}_t]$, 
	and $v_t \in [\underline{v}_t, \bar{v}_t]$ for all $t \geq 0$, where 
	$\underline{w}_t$, $\bar{w}_t$, $\underline{v}_t$, and $\bar{v}_t$ are known bounded signals.
\end{assum}

\begin{lem}\citep{efimov2013interval}\label{lemma:M interval}
	Given a vector $x \in \mathbb{R}^n$ satisfying $x \in [\underline{x}, \bar{x}]$ and a constant matrix $X \in \mathbb{R}^{m \times n}$,
	it holds that $X^{+} \underline{x} - X^{-} \bar{x}	\leq X x \leq X^{+} \bar{x} - X^{-} \underline{x}. $
\end{lem}

%\begin{lem}\citep{FarinaRinaldi00}\label{lemma:schur stable}
%	A positive definite matrix $X \in \mathbb{R}^{n \times n}$ is Schur stable if and only if there is a diagonal matrix $P \succ 0$ such that $X^\top P X - P \prec 0$.
%\end{lem}

\section{Interval Observer Design}\label{sec:observer}

The proposed interval observer needs the 
intervals for $x_0$, $w_t$, $v_t$, $f_\text{nn}(y^o_t,r_t)$, and $g(x(t))$. The 
intervals for $x_0$, $w_t$ and $v_t$ are known from Assumption 
\ref{assume:signal bounds}, while those for $f_\text{nn}(r_t, y^o_t)$ and $g(x(t))$ are computed below. 

\textbf{Interval of the NN Controller.}
Let the output $y_t^o$ satisfy $y_t^o \in [\underline{y^o}_t, \overline{y^o}_t]$.
It follows from Assumption \ref{assume:signal bounds} and Lemma \ref{lemma:M interval} that 
$\underline{y^o}_t = C_o ( C^{+} \underline{x}_t - C^{-} \overline{x}_t + \underline{v}_t)$ and  
$\overline{y^o}_t = C_o ( C^{+} \overline{x}_t - C^{-} \underline{x}_t + \bar{v}_t),
$
where $[\underline{x}_t, \overline{x}_t]$ is the interval of $x_t$ to be generated by the proposed interval observer. Suppose that the NN controller $f_\text{nn}(z_0)$ is bounded as
$f_\text{nn}(z_0) \in [\underline{f}_\text{nn}(t), \overline{f}_\text{nn}(t)]$ for the input region $z_0 \in [\underline{z}_0, \overline{z}_0]$, where $\underline{z}_0 = [r_t; \underline{y^o}_t]$ and $\overline{z}_0 = [r_t; \overline{y^o}_t]$. Two methods are described below to compute the NN interval $[\underline{f}_\text{nn}(t), \overline{f}_\text{nn}(t)]$.

The auxiliary network (AN) method described in \citep[Theorem 1]{Xiang21} computes the NN interval as follows:
\begin{subequations}\label{lemma:NN interval eq2}
	\begin{align}
		\label{eq:auxiliary NN lower}	
		&\left \{ 
		\begin{array}{l}
			\underline{z}_i = \phi(\underline{W}_i 
			\overline{z}_{i-1} + 
			\overline{W}_i \underline{z}_{i-1} + b_i), ~ 
			i \in [1, L-1] 
			\\
			\underline{f}_\text{nn}(t) = 
			\underline{W}_i 
			\overline{z}_{i-1} + 
			\overline{W}_i \underline{z}_{i-1} + b_i 
		\end{array}
		\right., \\
		%%%%%%%%%%%%%%%%%%%%
		\label{eq:auxiliary NN upper}	
		&\left \{ 
		\begin{array}{l}
			\overline{z}_i = \phi(\underline{W}_i 
			\underline{z}_{i-1} 
			+ \overline{W}_i \overline{z}_{i-1} + b_i), ~ i \in [1, L-1] 
			\\
			\overline{f}_\text{nn}(t) = 
			\underline{W}_i 
			\underline{z}_{i-1} 
			+ \overline{W}_i \overline{z}_{i-1} + b_i 
		\end{array}
		\right.	,
	\end{align}
\end{subequations}
where $\overline{W}_i = W_i^+$ and $\underline{W}_i = -W_i^-$.
This method ignores the coupling among neurons, resulting in a loose NN interval that will reduce the precision of the state trajectory intervals to be generated by the interval observer. 

In this paper, an optimisation method (OP) is proposed to compute a tighter (the best) NN interval by considering the coupling among all neurons.
Given the interval of $z_0$, the pre-activation bounds $\hat{l}_i$ and $\hat{u}_i$ of the $i$-th hidden layer, $i \in [1,L-1]$, can be computed using the interval arithmetic method as follows \citep{Liu+21}:
\begin{align}\label{eq:IA bounds}
	\begin{split}
		l_0 &= \underline{z}_0, ~u_0 = \overline{z}_0, \\ 
		\hat{l}_i &= \overline{W} l_{i-1} + \underline{W}_i u_{i-1} + b_i, ~ i \in [1,L-1],  \\
		\hat{u}_i &= \underline{W}_i l_{i-1} + \overline{W}_i u_{i-1} + b_i, ~ i \in [1,L-1],  \\
		l_i &= \phi(\hat{l}_i), ~ u_i = \phi(\hat{u}_i), ~ i \in [1,L-2].
	\end{split} 
\end{align}
The $j$-th \method{ReLU} neuron at the $i$-th hidden layer $z_{i,j} = \phi(W_i(j,:) z_{i-1} + b_i(j))$ is \textit{equivalently} represented by the following set of inequalities: 
\begin{align}\label{eq:MIP}
\begin{split}
	& \hat{z}_{i,j} = W_i(j,:) z_{i-1} + b_i(j), \\
	& z_{i,j} \geq \hat{z}_{i,j}, ~ z_{i,j} \geq 0, ~ z_{i,j} \leq \sigma_{i,j} \hat{u}_{i,j} , \\
	& z_{i,j} \leq \hat{z}_{i,j} - (1 - \sigma_{i,j}) \hat{l}_{i,j} , ~ \sigma_{i,j} \in \{0,1\},
\end{split}		
\end{align}
where $\hat{l}_{i,j}$ and $\hat{u}_{i,j}$ are the lower and upper pre-activation bounds of $\hat{z}_{i,j}$ computed in \eqref{eq:IA bounds}. 
The binary variable $\sigma_{i,j}$ indicates the status of the neuron: When $\sigma_{i,j} = 1$, the neuron is active and the above set of inequalities revert to be $z_{i,j} = \hat{z}_{i,j}$; when $\sigma_{i,j} = 0$, the neuron is inactive and the set of inequalities in \eqref{eq:MIP} revert to be $z_{i,j} = 0$.   

Based on \eqref{eq:MIP}, the $j$-th element of the lower bound $\underline{f}_\text{nn}(t)$ is solved from the following optimisation problem:
\begin{subequations}\label{OP:fnn min}
\begin{align}
& \underline{f}^j_\text{nn}(t) := \underset{\{z_i\}_{i=0}^{L-1}, \{\delta_i\}_{i=1}^{L-1}}{\min} W_L(j,:) z_{L-1} + b_L(j) \nonumber\\
& \hspace{-0.3cm} \text{s.t.}~~ 
\label{OP:fnn min const1}
 \underline{z}_0 \leq z_0 \leq \overline{z}_0, ~\hat{z}_i = W_i z_{i-1} + b_i, ~i \in [1,L-1],\\
\label{OP:fnn min const2}
& z_i \geq \hat{z}_i, ~ z_i \geq 0, ~ z_i \leq \sigma_i \odot \hat{u}_i, ~i \in [1,L-1],\\
\label{OP:fnn min const3}
& z_i \leq \hat{z}_i - (1 - \sigma_i) \odot \hat{l}_i, ~ \sigma_i \in \{0,1\}^{n_i}, ~i \in [1,L-1],
\end{align}
\end{subequations}
where $\odot$ is the elementwise product and $n_i$ is the number of neurons at the $i$-th hidden layer. 
The $j$-th element of $\overline{f}_\text{nn}(t)$ are solved from an optimisation problem in the same formulation as \eqref{OP:fnn min} but with $\min$ being replaced by $\max$. 
These optimisation problems are MILP problems which can be solved efficiently using off-the-shelf solvers like Gurobi \citep{gurobi}.

By using the computed NN interval $[\underline{f}_\text{nn}(t), \overline{f}_\text{nn}(t)]$, it follows from Assumption \ref{assume:signal bounds} and Lemma \ref{lemma:M interval} that
\begin{align}\label{eq:interval terms1}
	& B^{+} \underline{f}_\text{nn}(t) - B^{-} \overline{f}_\text{nn}(t)
	\leq 
	B f_\text{nn}(z_0) \nonumber\\
	& \leq 
	B^{+} \overline{f}_\text{nn}(t)  - B^{-} \underline{f}_\text{nn}(t).
\end{align}	

\textbf{Interval of the Nonlinear Function.}
The OVERT method in \citep{sidrane2022overt} is adapted to compute the interval $[\underline{g}(t), \overline{g}(t)]$ for $g(x_t)$. Its key idea is briefly introduced as below:
Let $g_i(x_t)$ be the $i$-th elementary nonlinear function of $g(x_t)$, where $i \in [1,s]$. Without generality, this paper assumes that $g_i(x_t)$ can be expressed by a conjunction of relations, where each relation is either an elementary function $e(\chi)$ of a scalar variable $\chi$ (e.g., $\sin(\chi)$, $\chi^2$, etc) or an algebraic operation (addition $+$ or subtraction $-$). The multiplication ($\times$) and division ($\div$) operations are converted into an exponential of a sum of logarithms, for which the details are referred to \citep[Algorithm 1]{sidrane2022overt}. Hence, the interval of $g(x_t)$ can be constructed by stacking the intervals of all elementary functions $g_i(x_t)$, $i \in [1,s]$. 
Let the scalar variable $\chi$ satisfy $\chi \in [\underline{\chi}, \overline{\chi}]$. 
The interval $[\underline{\chi}, \overline{\chi}]$ is divided into $h_i+1$ subintervals, where $h_i$ is the number of intermediate points between the two endpoints $\underline{\chi}$ and $\overline{\chi}$ to be selected.
We then form the upper (or lower) bound by connecting each two neighbouring points using either the secant line (tangent line for lower bound, resp.) if $g_i(\chi)$ is convex in that region, or the tangent line (secant line for lower bound, resp.) if $g_i(\chi)$ is concave. The obtained bounds are in the form of 
\begin{align}\label{eq:nonlinear bounds}
\overline{g}_i(\chi) = \sum_{j=0}^{h_i} a_{i,j} \varphi_{i,j}(\chi), ~
	\underline{g}_i(\chi) = \sum_{j=0}^{h_i} b_{i,j} \delta_{i,j}(\chi),		
\end{align}
where $\varphi_{i,j}(y_t^o)$ and $\delta_{i,j}(\chi)$ are linear functions, and $a_{i,j}$ and $b_{i,j}$ are constant coefficients. The optimal way of selecting the $h_i$ intermediate points is described in \cite{sidrane2022overt}. A higher value of $h_i$ gives a tighter interval but with a more complex expression. 

\begin{figure}[h]
	\vspace{-0.2cm}
	\centering
	\includegraphics[width=0.6\columnwidth]{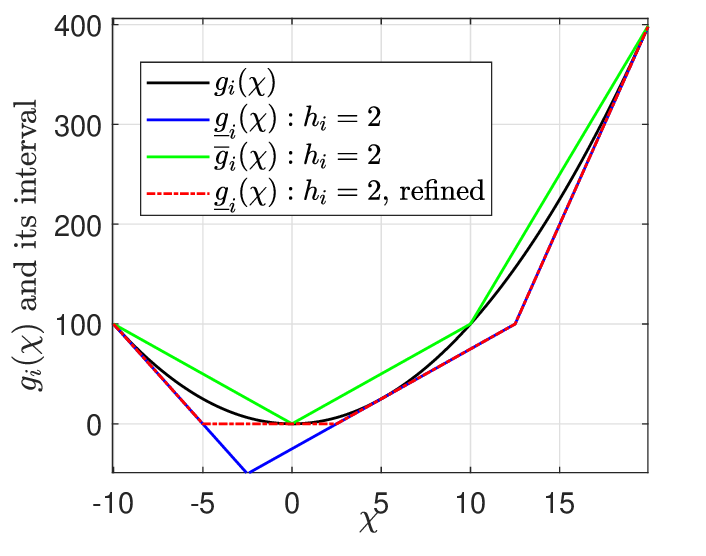} 
	\vspace{-4.5mm} 
	\caption{Intervals of $g_i(\chi) = \chi^2$ with and without refinement when $h_i = 2$ and $\chi \in [-10,20]$.}
	\label{fig0}
\end{figure}

Prior physical knowledge of the nonlinear function $g_i(\chi)$ can be used to refine the obtained interval. For example, in the simulation system in Section \ref{sec:simulation}, $g_i(\chi) = \chi^2$, where $\chi$ is the vehicle velocity. Hence, the lower bound of $g_i(\chi)$ satisfies $\underline{g}_i \geq 0$ and the lower bound in \eqref{eq:nonlinear bounds} is refined as $\underline{g}_i = \max (\underline{g}_i, 0)$. The refinement can help to obtain a tighter interval, as exemplified in Fig. \ref{fig0}.

The interval $[\underline{g}_i(x_t), \overline{g}_i(x_t)]$ for $g_i(x_t)$ under $x_t \in [\underline{x}_t, \overline{x}_t]$ can be derived using \eqref{eq:nonlinear bounds}, as in \citep[Algorithm 2]{sidrane2022overt}. But this interval depends on $x_t$, which is unavailable in the system \eqref{eq:sys dyn}. Instead, the state interval $x_t \in [\underline{x}_t, \overline{x}_t]$ will be provided at each time step by the proposed interval observer. This state interval can be used to compute $[\underline{g}_i(t), \overline{g}_i(t)]$ in real time, as described below: 

Given $x_t \in [\underline{x}_t, \overline{x}_t]$, the lower bound of $g_i(x_t)$ is redefined as $\underline{g}_i(t)$, by solving the following optimisation problem:
\begin{align}\label{OP:gx min}
 \underline{g}_i(t) := \quad & \underset{x_t, \Omega_i}{\min} \quad \Omega_i \nonumber \\
\text{s.t.} \quad & 
 \underline{x}_t \leq x_t \leq \overline{x}_t, ~
 \underline{g}_i(x_t) \leq \Omega_i \leq \overline{g}_i(x_t). 
\end{align}	 
The upper bound is redefined as $\overline{g}_i(t)$ and solved from an optimisation problem with the same form of \eqref{OP:gx min} but with $\min$ being replaced by $\max$. 

In some special cases, the nonlinear function $g_i(x_t)$ is monotonic in the region $x_t \in [\underline{x}_t, \overline{x}_t]$. One example is the simulation system in Section \ref{sec:simulation} where $g_i(x_t) = x_t^2$ and the scalar variable $x_t$ is the vehicle velocity satisfying $x_t \geq 0$. A graphic view of this example is shown in Fig. \ref{fig0} by regarding $x_t = \chi$ and $\chi \in [0,20]$. In such special cases, solving the optimisation problem \eqref{OP:gx min} is not needed. $\underline{g}_i(t)$ and $\overline{g}_i(t)$ can be directly computed using
\begin{equation}\label{OP:gx max}
\hspace{-0.28cm}\underline{g}_i(t) \!=\! \min \{ \underline{g}_i(\underline{x}_t),  \overline{g}_i(\underline{x}_t)\}, 
\overline{g}_i(t) \!=\! \max \{ \underline{g}_i(\overline{x}_t), ~ \overline{g}_i(\overline{x}_t)\}. \!\!\!\!
\end{equation}

By using the computed interval for $g(x_t)$ and Lemma \ref{lemma:M interval}, for $x_t \in [\underline{x}_t, \overline{x}_t]$, the interval of $F g(x_t)$ is derived as 
\begin{align}\label{eq:interval terms2}
	F^{+} \underline{g}(t) - F^{-} \overline{g}(t)
	\leq 
	F g(x_t) 
	\leq 
	F^{+} \overline{g}(t)  - F^{-} \underline{g}(t).
\end{align}

\textbf{Observer Design.}
For any design matrix $L_o \in \mathbb{R}^{n \times p}$, the system \eqref{eq:sys dyn} can be rewritten as
\begin{align}\label{eq:sys dyn2}
x_{t+1} = A_o x_t + B f_\text{nn} + F g(x_t)  + L_o y_t - L_o v_t + w_t.
\end{align}
where $A_o = A - L_o C$. 
Based on Assumption \ref{assume:signal bounds}, Lemma \ref{lemma:M interval}, \eqref{eq:interval terms1} and \eqref{eq:interval terms2}, the unknown terms $B f_\text{nn}(z_0)$, $F g(x_t)$, $L_o v_t$, and $w_t$ in~\eqref{eq:sys dyn2} are replaced with their intervals to formulate the interval observer:
\begin{subequations}\label{eq:interval observer}
\begin{align}
\overline{x}_{t+1} ={}& A_o \overline{x}_t + 
B^{+} \overline{f}_\text{nn}(t)  - B^{-} \underline{f}_\text{nn}(t) + F^{+} \overline{g}(t)  \nonumber\\
& - F^{-} \underline{g}(t) + L_o y_t - L_o^+ \underline{v}_t + L_o^- \overline{v}_t + \overline{w}_t, \\
\underline{x}_{t+1} ={}& A_o \underline{x}_t + 
B^{+} \underline{f}_\text{nn}(t) - B^{-} \overline{f}_\text{nn}(t) + F^{+} \underline{g}(t) \nonumber\\
&- F^{-} \overline{g}(t) + L_o y_t - L_o^+ \overline{v}_t + L_o^- \underline{v}_t + \underline{w}_t.
\end{align}
\end{subequations}
Define the interval width $\delta x_t = \overline{x}_t - 
\underline{x}_t$. Its dynamics are 
\begin{equation}\label{eq:interval width}
\delta x_{t+1} = A_o \delta x_t + D \xi_t,
\end{equation}
where $D = \left[|B|, |F|, |L_o|, I \right]$ and $\xi_t = [\delta f_\text{nn}; \delta g(t); \delta v_t; \delta w_t]$ is regarded as the disturbance with $\delta v_t = \overline{v}_t - \underline{v}_t$, $\delta w_t = 
\overline{w}_t - \underline{w}_t$, 
$\delta f_\text{nn}(t) = \overline{f}_\text{nn}(t) - \underline{f}_\text{nn}(t)$, and $\delta g(t) = \overline{g}(t) - \underline{g}(t)$.
%$|B| = B^+ + B^-$, $|F| = F^+ + F^-$, and $|L_o| = L_o^+ + L_o^-$.

The gain $L_o$ is designed to ensure that the interval observer \eqref{eq:interval observer} is sound, i.e., $\overline{x}_t \leq x_t \leq \underline{x}_t, \forall t \geq 0$, and the interval width $\delta x_t$ is robustly stable against the disturbance $\xi_t$. 
%The design is described in Theorem \ref{thm:Lo design}.

\begin{thm}\label{thm:Lo design}
For system \eqref{eq:sys dyn} under Assumption \ref{assume:signal bounds}, 
if there is a scalar $\rho$, a diagonal matrix $P$, and matrices $H_1, H_2$ such that the following optimisation problem is feasible:
\begin{subequations}\label{thm:op}	
\begin{align}
& \hspace{1.8cm}\underset{\rho, P, H_1, H_2}{\min} ~~ \rho	\nonumber\\
\label{thm:op const1}
&\text{s.t.}\hspace{0.5cm}  P \succ \mathbf{0}, ~  P A - (H_1 - H_2) C \geq \mathbf{0}, \\
\label{thm:op const3}
& \hspace{0.1cm}  \rho > 0, ~
\begin{bmatrix}
	P - I & \mathbf{0} & [P A - (H_1 - H_2) C]^\top  \\
	\star & \rho I & \hat{D}^\top \\
	\star & \star & P
\end{bmatrix}	\succ \mathbf{0},
\end{align}
\end{subequations}
where $\hat{D} = [P |B|,P |F|, H_1 + H_2, P]$, then \eqref{eq:interval observer} is a sound interval observer, i.e., $x_t \in [\underline{x}_t, 
\overline{x}_t] $, $\forall t \geq 0$, given that $x_0 \in [\underline{x}_0, 
\overline{x}_0]$. Moreover, the interval width $\delta x_t$ is robust against the disturbance vector $\xi_t$ with the $\mathrm{H}_\infty$ performance gain $\sqrt{\rho}$. The observer gain is computed as $L_o = L_o^{+} - L_o^{-}$ with $L_o^{+} = P^{-1} H_1$ and $L_o^{-} = P^{-1} H_2$. 
\end{thm}
\begin{pf}
Define the errors as $\overline{e}_t = \overline{x}_t - x_t$ and $\underline{e}_t = x_t - \underline{x}_t$. The error dynamics are derived as
\begin{equation}\label{thmpf:eq1}
\overline{e}_{t+1} = A_o \overline{e}_t + \overline{\Theta}_t, ~
\underline{e}_{t+1} = A_o \underline{e}_t + \underline{\Theta}_t,
\end{equation}
where 
$
\overline{\Theta}_t = B^+ (\overline{f}_\text{nn} - f_\text{nn}) 
+ B^- (f_\text{nn} - \underline{f}_\text{nn})
+ F^+ (\overline{g} - g) + F^- (g - \underline{g}) 
+ (\overline{w}_t - w_t)
+ L_o v_t - L_o^+ \underline{v}_t + L_o^- \overline{v}_t
$
and
$
\underline{\Theta}_t = B^+ (f_\text{nn} - \underline{f}_\text{nn}) 
+ B^- (\overline{f}_\text{nn} - f_\text{nn})  
+ F^+ (g - \underline{g}) + F^- (\overline{g} - g)  
+ (w_t - \underline{w}_t) 
- L_o v_t + L_o^+ \overline{v}_t - L_o^- \underline{v}_t
$.

Under Assumption \ref{assume:signal bounds}, the following inequalities hold:
\begin{align}
L_o v_t - L_o^{+} \underline{v}_t + L_o^{-} \overline{v}_t 
&= L_o^{+}(v_t - \underline{v}_t) + L_o^{-}(\overline{v}_t - v_t) \geq \mathbf{0}, \nonumber\\
- L_o v_t + L_o^+ \overline{v}_t - L_o^- \underline{v}_t 
&= L_o^{-} (v_t - \underline{v}_t) + L_o^{+} (\overline{v}_t - v_t) \geq \mathbf{0}. \nonumber
\end{align}
Subsequently, it holds that $\overline{\Theta}_t \geq \mathbf{0}$ and $\underline{\Theta}_t \geq \mathbf{0}$.
Further recalling that $\overline{e}_0, \underline{e}_0 \geq \mathbf{0}$ under Assumption \ref{assume:signal bounds}, it is seen from \eqref{thmpf:eq1} that $\overline{e}_t, \underline{e}_t \geq \mathbf{0}$, $\forall t \geq 0$, if $A_o$ is non-negative (i.e. all its elements are non-negative). 
Let $P \in \mathbb{R}^{n \times n} \succ \mathbf{0}$ be a diagonal matrix, then $A_o$ is non-negative if 
$
P A_o \geq \mathbf{0}.
$
Submitting $A_o = A - (L_o^{+} - L_o^{-}) C$ into the above inequality and introducing $H_1 = P L_o^{+}$ and $H_2 = P L_o^{-}$ yields \eqref{thm:op const1}. 

To ensure robust stability of $\delta x_t$ against the disturbance $\xi_t$, the Lyapunov function 
$V_t = \delta x_t^\top P \delta x_t$ is used. The following relation can be derived from \eqref{eq:interval width}: 
\begin{align}\label{thmpf:eq4}
\Delta V_t = V_{t+1} - V_t 
={}&
\zeta_t^\top 
\begin{bmatrix}
A_o^\top P A_o - P & A_o^\top P D \\
\star & D^\top P D
\end{bmatrix}
\zeta_t,
\end{align}
where $\zeta_t = [\delta x_t; \xi_t]$ and $D = \left[|B|, |F|, |L_o|, I \right]$.
Considers the performance index $J = \sum_{t=0}^{\infty} (\delta x_t^\top \delta x_t - \gamma^2 \xi_t^\top \xi_t + \Delta V_t )$ for a scalar $\gamma > 0$. If $J < 0$, the interval width dynamics \eqref{eq:interval width} satisfy the $\mathrm{H}_\infty$ performance $\sum_{t=0}^{\infty} \delta x_t^\top \delta x_t \leq \gamma^2 \sum_{t=0}^{\infty} \xi_t^\top \xi_t$. By using \eqref{thmpf:eq4}, a sufficient condition for $J < 0$ is $\delta x_t^\top \delta x_t - \gamma^2 \xi_t^\top \xi_t + \Delta V_t < 0$, i.e.,
\begin{equation}\label{thmpf:eq5}
\begin{bmatrix}
	A_o^\top P A_o - P + I & A_o^\top P D \\
	\star & D^\top P D - \gamma^2 I
\end{bmatrix}
\prec \mathbf{0}.
\end{equation}
Rearranging \eqref{thmpf:eq5} and applying Schur complement to it, with the introduction of new variables $H_1 = P L_o^{+}$, $H_2 = P L_o^{-}$, and $\rho = \gamma^2$, results in the constraint \eqref{thm:op const3}. \qed
\end{pf}

\section{Runtime Monitoring and Fault Detection}\label{sec:application}
%This section describes the application of the proposed interval observer \eqref{eq:interval observer} for monitoring real-time system safety and detecting potential faults. 

\textbf{Safety Monitoring.}
Given the safe state interval $\mathcal{X}_t$, it follows from Theorem \ref{thm:Lo design} that the $i$-th state $x_t^i$ is safe if 
\begin{equation}\label{eq:safe cond}
[\underline{x}_t^i, \overline{x}_t^i] \subseteq \mathcal{X}_t^i, ~i \in [1, n].	
\end{equation}
When this condition is false, safety of $x_t^i$ is undefined. It could be either $[\underline{x}_t^i, \overline{x}_t^i]$ is a too coarse outer-approximation 
of $x_t^i$ or the state $x_t^i$ is indeed unsafe. 
At time $t$, the observer \eqref{eq:interval observer} can also provide the one-step ahead predicted state interval $[\underline{x}_{t+1}, \overline{x}_{t+1}]$ and output interval $[\underline{y}_{t+1}, \overline{y}_{t+1}]$ as follows:
$\overline{y}_{t+1} = C^{+} \overline{x}_{t+1} - C^{-} \underline{x}_{t+1}  + \overline{v}_{t+1}
$,
$ 		
\underline{y}_{t+1} = C^{+} \underline{x}_{t+1} - C^{-}  \overline{x}_{t+1}  + \underline{v}_{t+1},
$
These one-step ahead predicted intervals can be used to raise alerts of potential unsafe operations in the future.
Denote the one-step ahead predicted intervals of $x_t$ and $y_t$ as 
$X_{\mathrm{pred},t+1} := [\underline{x}_{t+1}, 
\overline{x}_{t+1}]$ and $Y_{\mathrm{pred},t+1} := 
[\underline{y}_{t+1}, \overline{y}_{t+1}]$, respectively. 
Let the safe intervals of $x_{t+1}$ and $y_{t+1}$ be $\mathcal{X}_{t+1}$ and $\mathcal{Y}_{t+1}$, respectively. At time step $t+1$, the $i$-th state is safe if
\begin{equation}\label{eq:safe x}
X_{\mathrm{pred},t+1}^i \subseteq \mathcal{X}_{t+1}^i, ~ i \in [1, n],	
\end{equation}
and the $j$-th output is safe if
\begin{equation}\label{eq:safe y}
Y_{\mathrm{pred},t+1}^j \subseteq \mathcal{Y}_{t+1}^j, ~j \in [1, p].	
\end{equation} 
When these set inclusions are false, safety of the $i$-th state or $j$-th output is undefined.  
It could be either that the predicted intervals are too coarse or the real state or output are indeed unsafe. Nevertheless, the information of violation is still practically valuable for alerting unsafe system operations and triggering appropriate preventions.

\textbf{Fault Detection.} 
Consider the case when an actuator fault $f^{\text{a}}_t$ acting on the control signal $u_t$ as follows:
\begin{equation}\label{eq:actuator fault}
	u_t = f_\text{nn}(r_t,y^o_t) + f^{\text{a}}_t.
\end{equation}
By using the computed interval $[\underline{f}_\text{nn}(t), \overline{f}_\text{nn}(t)]$, there is a fault in the $i$-th control input channel if 
\begin{equation}\label{eq:FD actuator condition}
	u^i_t \notin [\underline{f}^i_\text{nn}(t), \overline{f}^i_\text{nn}(t)], ~i \in [1,m].
\end{equation}

The control input $u_t$ is also affected by faults occurring at the output $y^o_t$ because it is the input to the NN $f_\text{nn}$. 
In the presence of additive faults, the output $y^o_t$ is represented by
\begin{equation}\label{eq:sensor fault}
	y^o_t = C_o (C x_t + F_\mathrm{s} f^{\mathrm{s}}_t + v_t),
\end{equation} 
with the fault distribution matrix $F_\mathrm{s} \in \mathbb{R}^{p \times s}$, where 
$s \leq p$.
This paper focuses on sensor faults, but faults could also be from cyber attacks when some elements of $y^o_t$ are transmitted through communication networks. 
Presence of $f^{\mathrm{s}}_t$ can also be detected based on the condition \eqref{eq:actuator fault}. But this condition alone does not allow fault isolation because the fault effects due to $f^{\text{a}}_t$ and $f^{\mathrm{s}}_t$ are coupled.

Due to the nonlinear nature of the NN $f_\text{nn}$, it may happen that even a relative large output fault $f^{\mathrm{s}}_t$ results in just minor changes to $u_t$, making $u_t$ remain within the estimated interval and the fault ``invisible''. To address this issue, another detecting condition is introduced below:
The one-step ahead predicted interval for $y^o_t$ is  
$y^o_{t+1} \in [\underline{y}^o_{t+1}, \overline{y}^o_{t+1}]$, where $\underline{y}^o_{t+1} = C_o \underline{y}_{t+1}$ and $\overline{y}^o_{t+1} = C_o \overline{y}_{t+1}$. At time $t+1$, there is a fault in the $i$-th output $y^{o,i}_{t+1}$ 
if 
\begin{equation}\label{eq:output safe cond}
y^{o,i}_t \notin [\underline{y}^{o,i}_{t+1}, \overline{y}^{o,i}_{t+1}], ~i \in [1,\ell].
\end{equation}
If \eqref{eq:actuator fault} is false, but \eqref{eq:output safe cond} is true, then we are still able to know that there exists faults.

Sufficiency of the checking conditions \eqref{eq:safe cond}, \eqref{eq:safe x}, \eqref{eq:safe y}, \eqref{eq:FD actuator condition}, and \eqref{eq:output safe cond} relies on tightness of the generated intervals. Tighter intervals increase the capability of safety monitoring and fault detection. The tightness is affected by the priori known intervals for $\omega_t$ and $v_t$ in Assumption \ref{assume:signal bounds} and the intervals for $f_\text{nn}$ and $g(x_t)$ computed in Section \ref{sec:observer}. Those for $\omega_t$ and $v_t$ can be improved if a better knowledge of them is available. 
The proposed OP method has achieved the tightest interval for $f_\text{nn}$ with \method{ReLU} activations. For the nonlinear function $g(x_t)$, there may exist other ways to get a tighter interval, which is left for future research.

\section{Case Study}\label{sec:simulation}
Consider an adaptive cruise control (ACC) system:
\begin{subequations}\label{eq:simulation acc}
\begin{align}
\dot{x} ={}& A_c x + B_c u_\mathrm{e} + F_c g(x) + w, \\	
y ={}& C x + v, 
\end{align}
\end{subequations}
where $x = [p_\mathrm{l}; v_\mathrm{l}; a_\mathrm{l}; p_\mathrm{e}; v_\mathrm{e}; a_\mathrm{e}]$, $y = [p_\mathrm{e}; v_\mathrm{e}; h; \tilde{v}]$, $h = p_\mathrm{l} - p_\mathrm{e}$, $\tilde{v} = v_\mathrm{l} - v_\mathrm{e}$, 
$g(x) = [v_\mathrm{l}^2; v_\mathrm{e}^2]$, $w = [0; 0; 2 u_\mathrm{l}; 0; 0; 0]$,
and
\begin{align}
	A_c &= 
	\begin{bmatrix}
		0 & 1 & 0 & 0 & 0 & 0 \\
		0 & 0 & 1 & 0 & 0 & 0 \\
		0 & 0 & -2 & 0 & 0 & 0 \\
		0 & 0 & 0 & 0 & 1 & 0 \\
		0 & 0 & 0 & 0 & 0 & 1 \\
		0 & 0 & 0 & 0 & 0 & -2 
	\end{bmatrix}, 
	B_c = 
	\begin{bmatrix}
		0 \\ 0 \\ 0 \\ 0 \\ 0 \\ 2
	\end{bmatrix}, 
	F_c = 
	\begin{bmatrix}
		0 & 0 \\ 0 & 0 \\ -\mu & 0 \\ 0 & 0 \\ 0 & 0 \\ 0 & -\mu
	\end{bmatrix}. \nonumber 
\end{align}
The variables $p$, $v$, $a$, and $u$ are the position, velocity, acceleration, and control command of the lead (with subscript $\mathrm{l}$) and ego vehicles (with subscript $\mathrm{e}$), respectively. 
$\mu$ is the friction parameter. 
Discretising \eqref{eq:simulation acc} with sampling time $t_s = 0.1 ~\mathrm{s}$ gives a system in the form of \eqref{eq:sys dyn}. 

Define the safe relative vehicular distance as $d_\mathrm{safe} = v_\mathrm{e} t_\mathrm{gap} + d_\mathrm{still}$, where $t_\mathrm{gap}$ is the time headway and $d_\mathrm{still}$ is the standstill distance. To achieve safe car following, the ACC controller $u_\text{e}$ is designed for the ego vehicle to achieve two objectives: When $h \geq d_\mathrm{safe}$, $u_\mathrm{e}$ takes the speed control mode to maintain the ego vehicle at the driver-set speed $v_\mathrm{set}$; When $h < d_\mathrm{safe}$, $u_\mathrm{e}$ takes the spacing control mode to ensure $h = d_\mathrm{safe}$. A 4-layer feedforward NN controller $u_\text{e} = f_\mathrm{nn}(r,y^o)$, each hidden layer having 10 \method{ReLU} neurons, is trained and implemented, where 
$
r = [v_\mathrm{set}; t_\mathrm{gap}]$ and
$y^o  = [h; \tilde{v}; v_\mathrm{e}] = C_o y$ with $C_o = [\mathbf{0}_{3 \times 1}, I_3]$. 
The simulation parameters are: $\mu \!=\! 0.0001$, $t_\mathrm{gap} 
= 1.4 ~\mathrm{s}$, $d_\mathrm{still} \!=\! 10 ~\mathrm{m}$, 
$v_\mathrm{set} \!=\! 30 ~\mathrm{m/s}$, $a_\mathrm{min} \!=\! -3 
~\mathrm{m/s^2}$, $a_\mathrm{max} \!=\! 2 ~\mathrm{m/s^2}$. $v$ is 
a 
zero-mean white noise with $|v|  \!\leq\! 0.001$. 
The initial 
state is $x_0  \!=\! [50; 20; 0; 10; 20; 0]$. The 
signal 
bounds in Assumption \ref{assume:signal bounds} are set as $\underline{x}_0 \!=\! [49; 19; -1; 9; 19;-1]$, $\overline{x}_0 \!=\! [51; 21; 1; 11; 21;1]$, $\underline{v}_t \!=\! -0.001$, and 
$\overline{v}_t \!=\! 0.001$.
Comparison is made for the interval observers using the auxiliary network (AN) or optimisation (OP) methods to compute the interval of $f_\text{nn}$. At each time step, the MILP problems are solved using Gurobi and their total computation time is 0.006 s in average. Vehicle reversing is not considered in the simulation, thus each element of the nonlinear function $g(x) = [v_\mathrm{l}^2; v_\mathrm{e}^2]$ is monotonic and the interval for $g(x)$ is computed using \eqref{OP:gx max}.

\begin{figure}[h]
		\vspace{-2.6mm} 
	\centering
	\includegraphics[width=0.7\columnwidth]{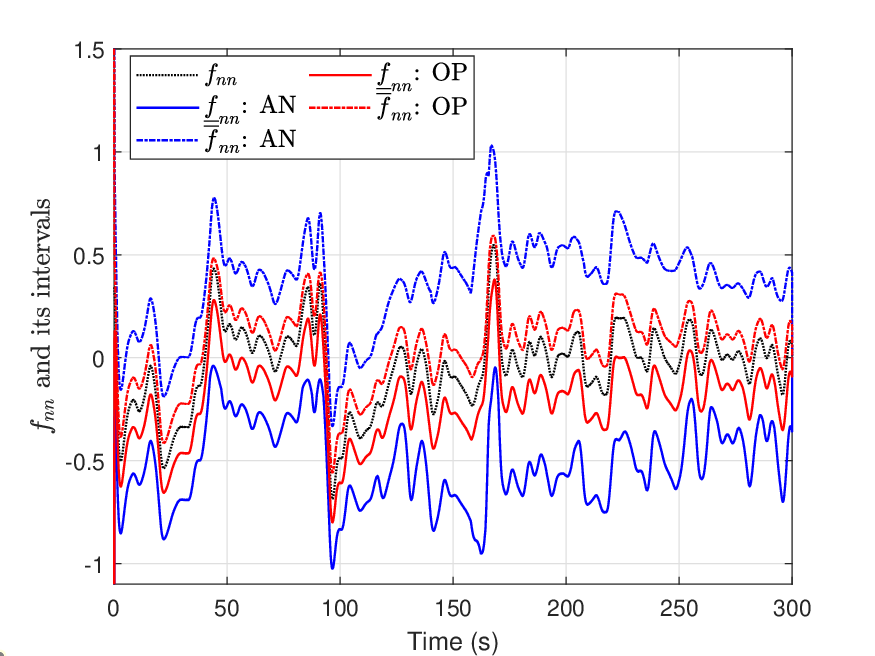} 
	\vspace{-3mm} 
	\caption{NN controller $u_\text{e} = f_\mathrm{nn}(r,y^o)$ and its intervals.}
	\label{fig1}
\end{figure}

\begin{figure}[t]
	\vspace{-2mm} 
	\centering
	\includegraphics[width=0.7\columnwidth]{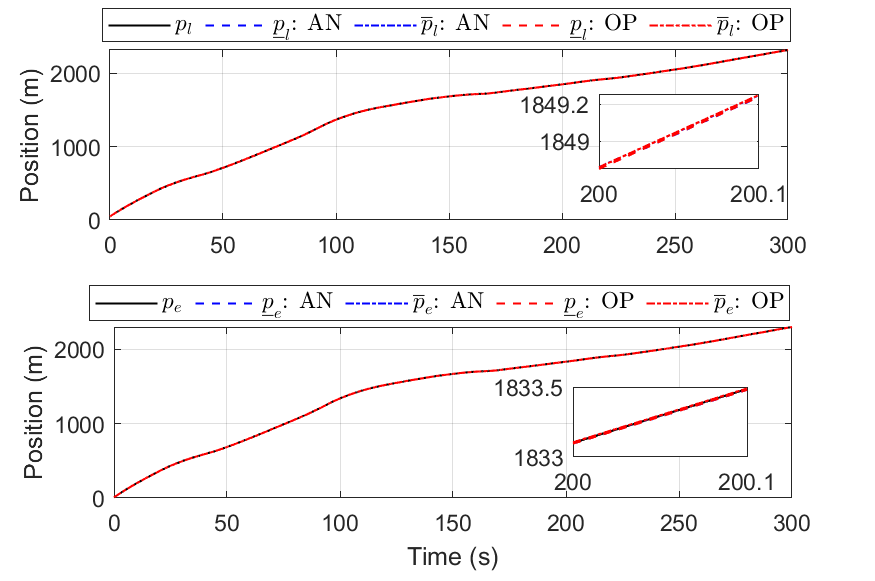} 
	\vspace{-3mm} 
	\caption{Vehicle positions ($p_\mathrm{l}$ and $p_\mathrm{e}$) and their intervals.}
	\label{fig2}
\end{figure}

\begin{figure}[t]
	\vspace{-2mm} 
	\centering
	\includegraphics[width=0.7\columnwidth]{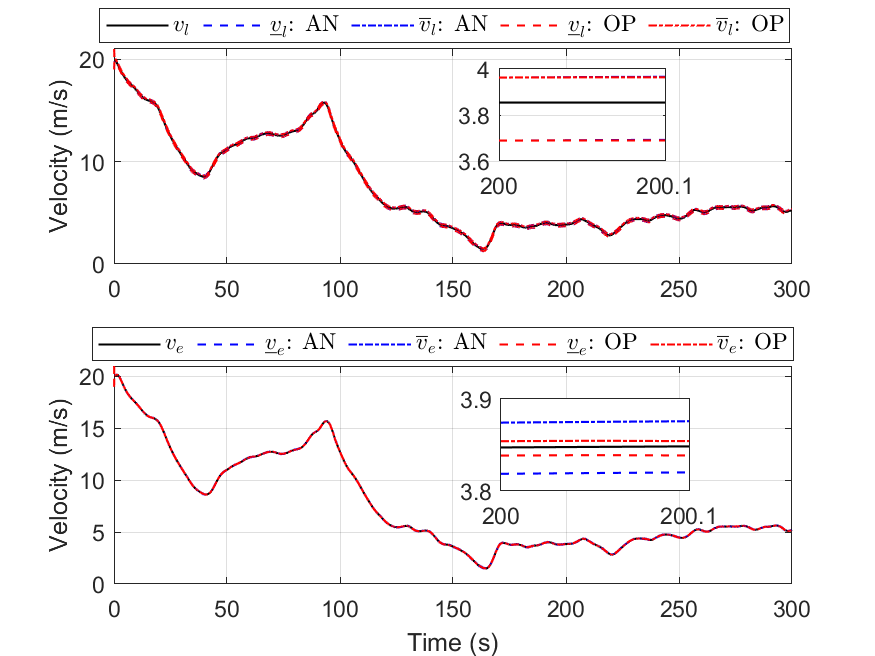} 
	\vspace{-3mm} 
	\caption{Vehicle velocities ($v_\mathrm{l}$ and $v_\mathrm{e}$) and their intervals.}
	\label{fig4}
\end{figure}

\begin{figure}[t]
	\vspace{-3mm} 
	\centering
	\includegraphics[width=0.7\columnwidth]{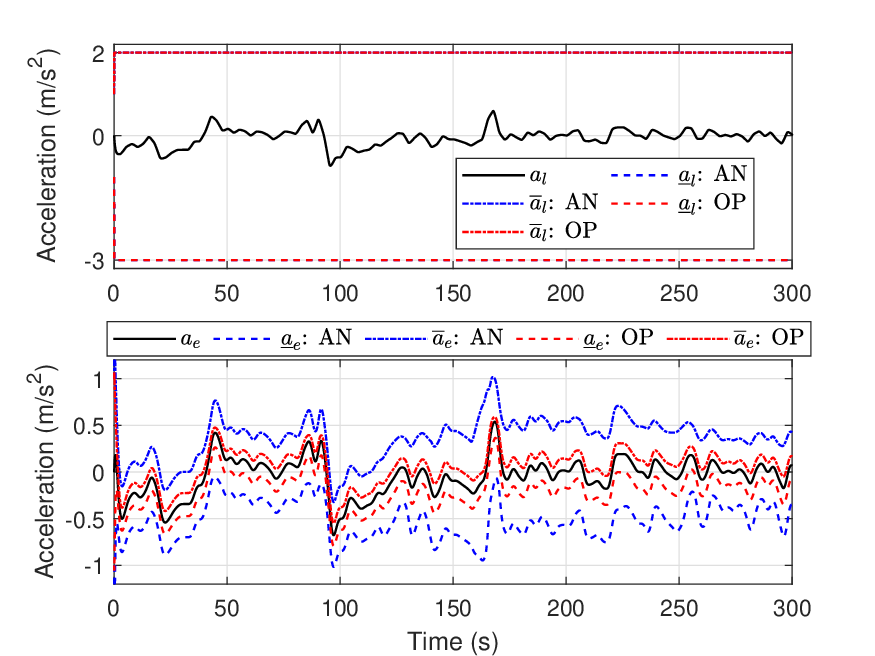} 
	\vspace{-3.5mm} 
	\caption{Vehicle accelerations ($a_\mathrm{l}$ and $a_\mathrm{e}$) and their intervals.}
	\label{fig5}
\end{figure}

\textbf{Case 1:} The results in Figs. \ref{fig1} - \ref{fig5} demonstrate efficacy of the proposed observer in generating sound and robust runtime intervals for the NN controller, vehicle positions, velocities, and accelerations. The results also demonstrate that the intervals based on the OP method are much tighter than those using the AN method, including $f_\text{nn}$ in Fig. \ref{fig1}, $v_\text{e}$ in Fig. \ref{fig4}, and $a_\text{e}$ in Fig. \ref{fig5}. The two methods result in similarly tight intervals for the other vehicle states.

\begin{figure}[h]
	\vspace{-2mm}
	\centering
	\includegraphics[width=0.7\columnwidth]{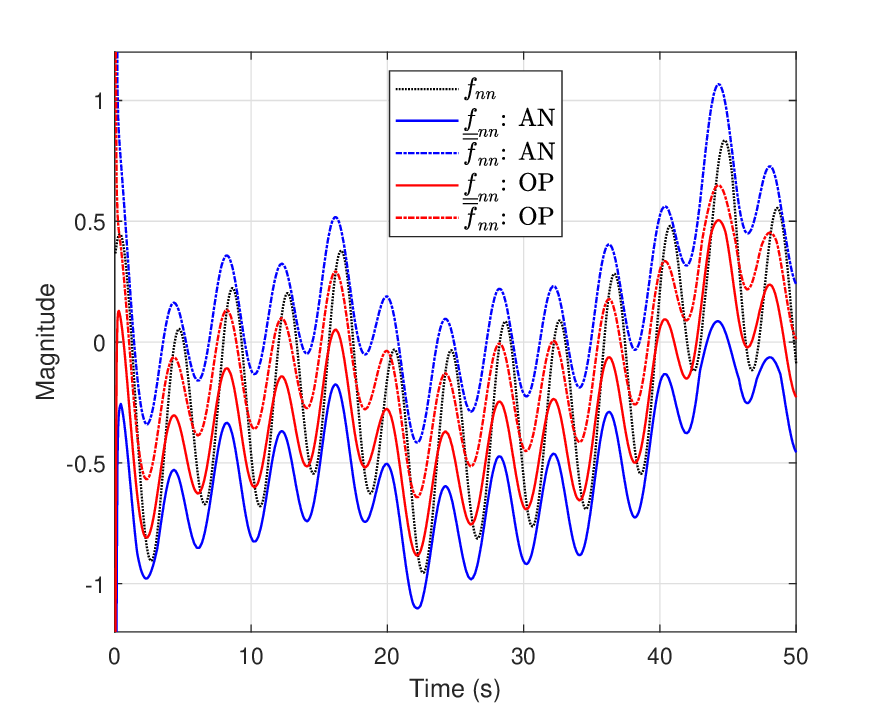} 
	\vspace{-3mm} 
	\caption{Detection of actuator fault $f_a(t) = 0.3 \sin(0.5 \pi t_s t)$.}
	\label{fig6}
\end{figure}

\textbf{Case 2:} 
Consider the case when there is a bias actuator fault $f_a(t) = 0.3 \sin(0.5 \pi t_s t), t \geq 0$. 
Fig. \ref{fig6} shows that if the interval of $f_\text{nn}$ is generated by the OP method, then applying the condition \eqref{eq:FD actuator condition} detects the fault. If the interval is generated by the AN method, the fault is not detected. This showcases that the OP method is advantageous over the AN method in actuator fault detection. 

\begin{figure}[h]
	\centering
	\includegraphics[width=0.7\columnwidth]{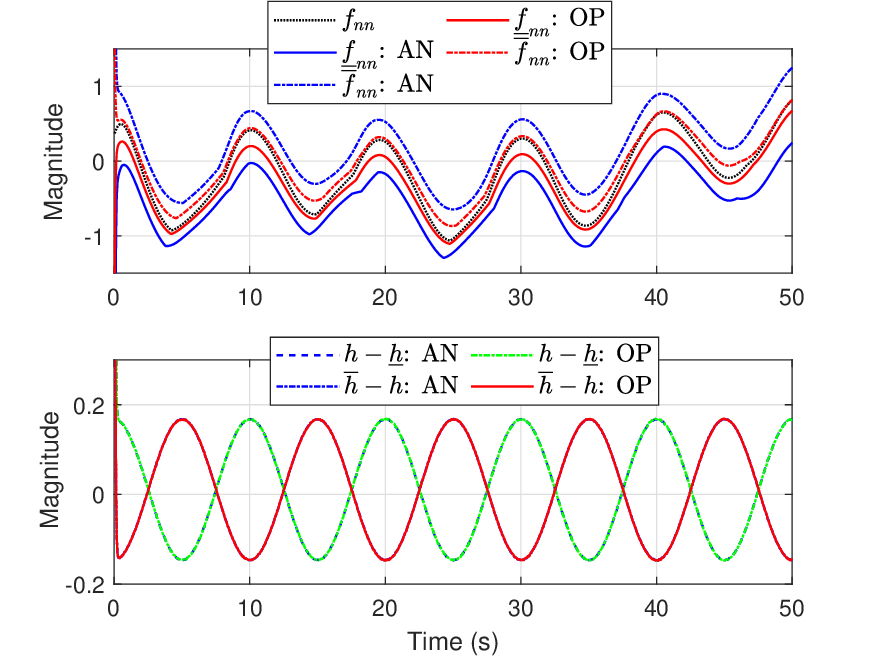} 
	\vspace{-3mm} 
	\caption{Detection of fault $f_s(t) = 5\sin(0.2 \pi t_s t)$  on $p_\text{l}$.}
	\label{fig7}
\end{figure}

\begin{figure}[h]
	\vspace{-1.5mm} 
	\centering
	\includegraphics[width=0.7\columnwidth]{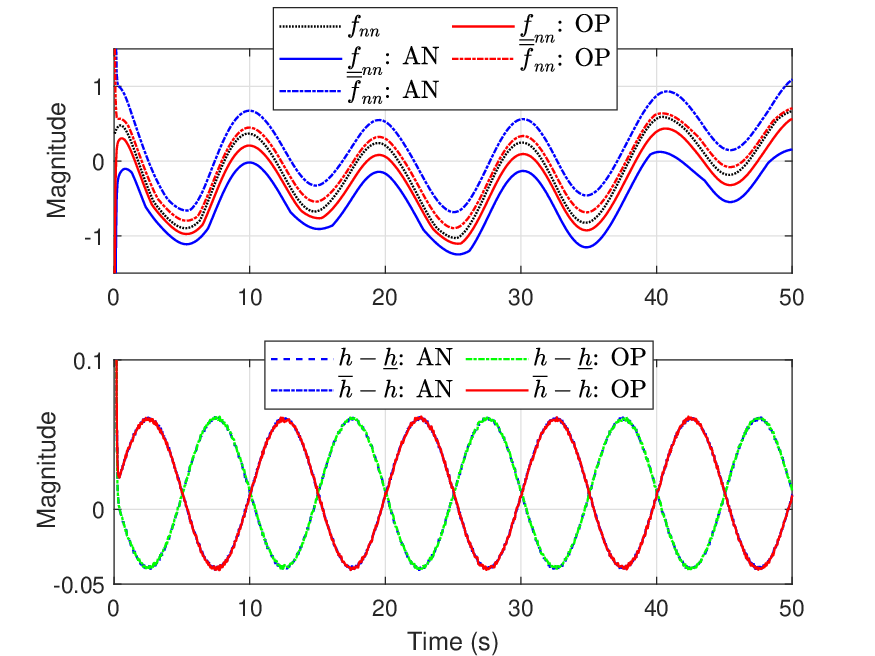} 
	\vspace{-3mm} 
	\caption{Detection of fault $f_s(t) = \sin(0.2 \pi t_s t)$ on $v_\text{l}$.}
	\label{fig8}
\end{figure}

Suppose the lead vehicle sends its real-time position $p_\mathrm{l}$ and velocity $v_\mathrm{l}$ to the ego vehicle through a wireless communication network. These data are used to construct $y^o_t$ for generating the NN control law $f_\text{nn}$. 
Consider the case when there is a relative large fault $f_s(t) = 5\sin(0.2 \pi t_s t)$ on the data $p_\text{l}$ or  
$f_s(t) = \sin(0.2 \pi t_s t)$ on the data $v_\text{l}$. These faults may be due to offsets in the lead vehicle's sensors \citep{LanZhao20r} or cyber attacks in the communication network \citep{PetitShladover14}. 
The top subplots in Fig. \ref{fig7} and Fig. \ref{fig8} show that using the interval of $f_\text{nn}$ from either the AN or OP methods are unable to detect the faults, meaning that the condition \eqref{eq:FD actuator condition} alone is not sufficient. The bottom subplots in Fig. \ref{fig7} and Fig. \ref{fig8} show that these faults can be effectively detected by both methods using the interval $[\underline{h}, \overline{h}]$ of the output $h$ based on the condition \eqref{eq:output safe cond}.
If no faults present, the interval observer always generates the interval satisfying $\underline{h} \leq h \leq \overline{h}$, i.e., $h - \underline{h} \geq 0$ and $\overline{h} - h \geq 0$. However, these two inequalities are violated as shown in the bottom subplots in Fig. \ref{fig7} and Fig. \ref{fig8}, indicating the existence of faults in the system. This confirms the benefits of combining the two conditions \eqref{eq:FD actuator condition} and \eqref{eq:output safe cond}, as discussed in Section \ref{sec:application}.

%%%%%%%%%%%%%%%%%%%%%%%

\section{Conclusions}\label{sec:conclusion}

A robust interval observer is proposed to generate a sound and tight state interval for nonlinear systems controlled by a deep NN. 
The obtained interval is applied to monitor the runtime system safety and  
detect faults in the actuators and sensors.  
Simulation results of the ACC system 
demonstrate effectiveness of the interval observer and its 
advantages over the existing design.
As for future research, appropriate remedial actions will be proposed to ensure safe 
operation of NN-controlled systems based on the monitoring and detection results.

\bibliographystyle{ifacconf}
\bibliography{reference}

%\appendix
%\input{sections/appendix}

\end{document}